\documentclass[pre,reprint,superscripts]{revtex4-2}
\usepackage{amsmath}
\usepackage{amssymb}
\usepackage{graphicx}
\usepackage[usenames,dvipsnames]{color}
\usepackage[hidelinks]{hyperref}

\def\g{\Gamma}
\newcommand{\av}[1]{\left\langle {#1 }\right\rangle}

\newcommand{\new}[1]{{\color{black} #1}}

\begin{document}

\title{Exact finite-dimensional description for networks of globally coupled spiking neurons}
\author{Bastian Pietras}
\email{bastian.pietras@upf.edu}
\affiliation{Department of Information and Communication Technologies, Universitat Pompeu Fabra, T\`anger 122-140, 08018, Barcelona, Spain}
\author{Rok Cestnik}
\email{rokcestn@uni-potsdam.de}
\affiliation{Department of Physics and Astronomy, University of Potsdam, Karl-Liebknecht-Str. 24/25, 14476, Potsdam-Golm, Germany}
\author{Arkady Pikovsky}
\email{pikovsky@uni-potsdam.de}
\affiliation{Department of Physics and Astronomy, University of Potsdam, Karl-Liebknecht-Str. 24/25, 14476, Potsdam-Golm, Germany}
\date{\today}

\begin{abstract}
We consider large networks of globally coupled spiking neurons and derive an exact low-dimensional description 
of their collective dynamics in the thermodynamic limit. Individual neurons are described by the Ermentrout-Kopell canonical model that can be excitable or tonically spiking, and interact with other neurons via pulses. 
Utilizing the equivalence of the quadratic integrate-and-fire and the theta neuron formulations, we first derive the 
dynamical equations in terms of the Kuramoto-Daido order parameters (Fourier modes of the phase distribution) 
and relate them to two biophysically relevant macroscopic observables, the firing rate and the mean voltage. 
For neurons driven by Cauchy white noise or for Cauchy-Lorentz distributed input currents, we adapt the 
results by Cestnik and Pikovsky [\new{Chaos {\bf 32}, 113126 (2022)}] and show that for arbitrary initial conditions the collective dynamics 
reduces to six dimensions. We also prove that in this case the dynamics asymptotically converges to a two-dimensional 
invariant manifold first discovered by Ott and Antonsen. For identical, noise-free neurons, the dynamics reduces to 
three dimensions, becoming equivalent to the Watanabe-Strogatz description. 
We illustrate the exact six-dimensional dynamics outside the invariant manifold by calculating nontrivial basins
of different asymptotic regimes in a bistable situation.
\end{abstract}
\maketitle


\section{Introduction}

Synchronization phenomena in ensembles of coupled oscillators is an active field
of interdisciplinary research with numerous applications in physics, engineering, and life sciences (see, e.g., 
books and reviews~\cite{Strogatz-03,Pikovsky-Rosenblum-Kurths-01,Acebron-etal-05,Pikovsky-Rosenblum-15}).
One important area of application is neuroscience, where synchronization of neurons is crucial
for understanding brain functioning~\cite{Buzsaki-06,Izhikevich-07}. While for many systems in physics and engineering the basic
equations for the oscillator dynamics can be formulated in the form of the Kuramoto model and its modifications~\cite{Acebron-etal-05},
neural models usually follow a different formulation that takes into account specific properties of spiking neurons and their interaction.
Nevertheless, studies of collective effects in large neural populations have profited enormously from the analogy of particular models of neural dynamics to the Kuramoto model \new{\cite{clusella2022kuramoto}}. 
Notably, a breakthrough in the
description of the Kuramoto-type dynamics by Ott and Antonsen~\cite{Ott-Antonsen-08} has recently been transferred to the realm of spiking neuron networks~\cite{Luke-Barreto-So-13,So-Luke-Barreto-14,Laing-14,montbrio_pazo_roxin_2015,laing_2015,clusella_montbrio_2022}, see also the reviews~\cite{ashwin_coombes_nicks_2016,bick_goodfellow_laing_martens_2020}.

The main finding of Ott and Antonsen is the existence of an invariant two-dimensional manifold, the so-called Ott-Antonsen (OA) manifold, corresponding to a
wrapped Cauchy distribution of the oscillators' phases, which allows for a formulation of exact closed
equations of motion for the global order parameter. In terms of the neural dynamics, these equations
correspond to closed equations for the parameters characterizing populations of neurons, such as the firing rate and mean voltage \new{\cite{montbrio_pazo_roxin_2015}}.
While the OA manifold is generally assumed to be attractive for systems with quenched and/or noisy inhomogeneity, and thus
describes asymptotic in time attractors, generally initial conditions lie outside the OA manifold and the corresponding 
transients are not captured by the OA equations. Recently, two of us developed an exact six-dimensional 
description for the evolution of Kuramoto-type oscillator populations outside of the OA manifold~\cite{cestnik_pikovsky_2022a}. 
The goal of this paper is to extend
this approach to networks of \new{globally coupled spiking neurons, 
whose microscopic dynamics are given by the quadratic integrate-and-fire (QIF) or the theta-neuron model,
\begin{gather*}
\dot v_k = v_k^2 - g v_k + I(t) + \Gamma I_k(t), \quad k=1,\dots,N,\\
\dot \theta_k = 1 - \cos \theta_k - g \sin \theta_k + (1+\cos\theta_k) \big[ I(t) + \Gamma I_k(t)\big],
\end{gather*}
with global recurrent input $I(t)$ and individual inputs $I_k(t)$ that include quenched and noisy inhomogeneity (for a detailed discussion of these equations see Section~\ref{sec:gcsn} below).
In~\cite{clusella_montbrio_2022} it was demonstrated that the collective dynamics of QIF neurons driven by independent Cauchy white noise is described by a mean-field model on the OA manifold that is identical to that for QIF neurons driven by time-independent Cauchy-Lorentz distributed inputs.
Here, we will go beyond the OA theory and demonstrate that the collective dynamics in the thermodynamic limit $N\to\infty$ are exactly described by system~\eqref{eq:ndv},
\begin{equation*}
	\dot{\Phi} = i \Phi^2 -g\Phi - i I(t) + \Gamma , \ \ \
	\dot{\lambda} = 2i\Phi\lambda - g\lambda, \ \ \ \dot{\sigma} = i\lambda,
\end{equation*}
for three complex collective variables $\Phi,\lambda,\sigma$ (Section~\ref{sec:fdr} below is devoted to the
derivation of this principal result of our work).
These variables readily allow for extracting the dynamics of the population firing rate $R(t)$ and of the mean voltage $V(t)$ through a simple relation, Eq.~\eqref{eq:RV_vars}.}
With these exact six-dimensional equations one can properly
describe the transient dynamics outside the OA manifold, and in particular find exact basins of attraction of different
asymptotic dynamical regimes (which lie on the OA manifold) for arbitrary perturbations.

The paper is organized as follows. In Section \ref{sec:gcsn} we introduce \new{the two basic equivalent spiking neuron models,}
the theta-neuron and the QIF neuron, and derive equations for the order parameters in the thermodynamic
limit in presence of inhomogeneity and noise. In Section \ref{sec:obs} we discuss the mean field observables and the coupling terms.
Section \ref{sec:fdr} contains our main findings. Here we derive the six-dimensional reduction of the dynamics, and show how the initial states of the neurons translate into initial conditions for the low-dimensional dynamics. 
We also discuss the case of identical units (in the absence of noise and heterogeneity).
In Section \ref{sec:oa} we discuss the relation to the OA theory, and in particular we demonstrate how the stability
of the OA manifold manifests itself within our formalism. In Section~\ref{sec:nex} we present several examples of application
of our approach to the dynamics off the OA manifold, including numerical simulations of finite ensembles of spiking neurons. We discuss our results in Section~\ref{sec:disc}.

\section{Globally coupled spiking neurons}
\label{sec:gcsn}
\subsection{Theta-neuron and QIF formulations} 
In this paper we consider globally coupled spiking neurons described by the Ermentrout-Kopell canonical 
model~\cite{Ermentrout_Kopell_86}. There are two equivalent formulations of this model:
one (theta-neuron, TN) uses a continuous phase-type variable $\theta(t)$; the other (quadratic integrate-and-fire neuron, QIF) uses a variable $v$ which is roughly interpreted as a membrane potential (for
mathematical simplicity one allows $v$ to attain infinite values, thus it is not a real membrane potential of a neuron). 
We present both versions in parallel, cf.~\cite{ermentrout_1996type,Luke-Barreto-So-13,So-Luke-Barreto-14,Laing-14,montbrio_pazo_roxin_2015,laing_2015}. 

A QIF neuron is described by the equation
\begin{equation}
\dot{v} = v^2 -gv + I(t) \;,
\label{eq:QIF1}
\end{equation}
where $I(t)$ is the input current that together with the gap junction coupling parameter $g$ determines the
dynamics of the neuron. Because of the quadratic nonlinearity, the voltage variable $v$ may reach infinity in a finite time, at which point it is reset to $-\infty$, and this event is interpreted as a spike.

To obtain the equation for a theta-neuron, one makes a transformation $v(t) = \tan(\theta(t)/2)$,
so that Eq.~\eqref{eq:QIF1} takes on the form
\begin{equation}
\dot{\theta}= 1-\cos \theta -g \sin \theta + (1+\cos\theta) I(t)=\omega+\frac{he^{-i\theta}-h^*e^{i\theta}}{i}\; .
\label{eq:TN1}
\end{equation}
Here, the spike occurs when the phase variable $\theta$ passes through $\pi$ with positive speed. 
Hence, the spike form can be modelled with a
function of the variable $\theta$ having a peak at $\theta\approx \pi$. We have introduced parameters
\begin{equation}
\omega(t)=1+I(t),\qquad h(t)=\frac{g+i[I(t)-1]}{2}\;,
\label{eq:w_h}
\end{equation}
because they allow to interpret Eq.~\eqref{eq:TN1} as a Kuramoto-Sakaguchi model for 
phase oscillators~\cite{kuramoto_model,Sakaguchi-Kuramoto-86,Acebron-etal-05}, with natural frequency $\omega$ and driving field $h$. 

The QIF dynamics \eqref{eq:QIF1} represents a discontinuous dynamical system that may encompass 
real conceptual and mathematical difficulties due to the instantaneous reset whenever a neurons crosses the 
threshold at infinity~\cite{cessac_vieville_2008,kevrekidis_siettos_kevrekidis_2017}. For this reason, it may be 
advantageous to consider the TN dynamics \eqref{eq:TN1}, which is bounded and smooth, circumvents 
the fire-and-reset discontinuity, and can \new{be treated as a smooth dynamical system}. 

\subsection{Population of identical neurons in the thermodynamic limit}
We now consider a population of identical neurons, which means that for all of them the values of $g,I(t)$
are the same. In the thermodynamic limit of an infinite number of units, a proper description
is via the probability density. In the TN formulation, the equation of probability conservation
for this density $P(\theta,t)$ reads
\new{
\begin{equation}
\frac{\partial P}{\partial t}+\frac{\partial }{\partial \theta}
\Big[ \big[ \omega+\frac{1}{i}(h e^{-i\theta}-h^*e^{i\theta}) \big] P(\theta,t) \Big]=0\;.
\label{eq:densth}
\end{equation}
}
In the QIF formulation, one has density $W(v,t)$ which is related to $P(\theta,t)$ as
\begin{equation}
W(v)=\frac{2P(2\arctan(v))}{1+v^2}\;, \quad P(\theta) = \frac{W(\tan(\theta/2))}{1+\cos(\theta)}\;,
\label{eq:ptow}
\end{equation}
(we have omitted the dependence on time for readability). 
This density obeys the probability conservation equation
\new{\[
\frac{\partial W}{\partial t}+\frac{\partial }{\partial v}
\Big[\big[ v^2-g v+I\big] W(v,t)\Big]=0\;,
\]}
which should be equipped with the \new{consistency} condition (outgoing flux at ``threshold'' $v=\infty$
should be equal to the incoming flux at ``reset'' $v=-\infty$).

The relation to the Kuramoto model mentioned above suggests the use of the Kuramoto-Daido
order parameters
\begin{subequations}
\begin{gather}
z_n(t)=\int_0^{2\pi}d\theta P(\theta,t)e^{i n\theta}=\int_{-\infty}^{\infty}d v\;W(v,t) \left(\frac{1+iv}{1-iv}\right)^n\;,\label{eq:KD_def}\\ 
P(\theta,t)=\new{\frac{1}{2\pi}}\sum\limits_{n=-\infty}^\infty z_n(t) e^{-in\theta}\;,
\end{gather}
\label{eq:ordpar}
\end{subequations}
to describe the evolution, which now reduces to an infinite system
\begin{equation}
\dot z_n=n[i\omega z_n+h z_{n-1}-h^*z_{n+1}]\;.
\label{eq:zident}
\end{equation}
This system is a Fourier representation of Eq.~\eqref{eq:densth}. \new{Due to the symmetry $z_{-n} = z_n^*$ that follows from definition~\eqref{eq:KD_def} of the Kuramoto-Daido order parameters, it suffices to consider only  positive indices $n>0$.} Since the probability density is normalized, we have $z_0=1$.

The order parameter representation of the probability density deserves a more detailed discussion.
This representation is 
straightforward in the TN formulation, because there it is just the Fourier
series representation of a periodic function. Correspondingly, the order parameter description is adequate if the 
probability density $P(\theta,t)$ is smooth enough \cite{zygmund2002trigonometric}. For the QIF model,
the distribution density $W(v,t)$ is defined on the line $-\infty<v<\infty$, and one may expect that
a discrete series representation would be insufficient. However, because we define the density $W$ via
the transformation \eqref{eq:ptow}, smoothness of $P$ implies a restriction on the possible behavior of $W$ so that the series representation is valid for the QIF model as well. 

\subsection{Effect of Cauchy noise}
Here we generalize the deterministic model above and consider spiking QIF neurons~\eqref{eq:QIF1} subject to
additive independent, identically distributed Cauchy noise $\xi(t)$ (cf.~\cite{Toenjes-Pikovsky-20}); 
a particular case of this generalization is also considered in~\cite{clusella_montbrio_2022}. 
The TN dynamics~\eqref{eq:TN1} now reads   
\begin{equation}
\dot\theta_k=1-\cos\theta_k-g\sin\theta_k+(1+\cos\theta_k)\big[I(t)+\gamma\xi_k(t)\big]\;,
\label{eq:thcn}
\end{equation}
where $\gamma$ is the noise intensity, and we assign index $k$ to neurons to stress the independence
of the noise terms $\xi_k$ and $\xi_m$ for $k\neq m$.
Then instead of Eq.~\eqref{eq:densth} we obtain a generalized Fokker-Planck equation (in the Stratonovich interpretation)
\new{
\begin{equation}
\begin{aligned}
\frac{\partial P}{\partial t}+\frac{\partial }{\partial \theta}
\Big[ \big[ 1-\cos\theta-g&\sin\theta+(1+\cos\theta)I \big] P \Big]=\\
=\, &\gamma
\frac{\partial }{\partial |\theta|}(1+\cos\theta) P(\theta,t)\;,
\end{aligned}
\label{eq:gfpth}
\end{equation}
}
where in the Fourier space the operator $\frac{\partial }{\partial |\theta|}$ acts as $\frac{\partial }{\partial |\theta|}e^{ik\theta}=-|k|e^{ik\theta}$.
If we insert the Fourier series \eqref{eq:ordpar}, then 
\new{
\begin{gather*}
\frac{\partial }{\partial |\theta|}(1+\cos\theta) P(\theta,t)= \sum\limits_{n=-\infty}^\infty -z_n \kappa_n(\theta), \quad \text{where}\\
\kappa_n(\theta) = |n|e^{-in\theta}+\frac{|n+1|}{2}e^{-i(n+1)\theta}+\frac{|n-1|}{2}e^{-i(n-1)\theta}.
\end{gather*}
}
This leads to the following modification of Eqs.~\eqref{eq:zident} for the order parameters $z_n, n\geq 1$,
\begin{equation}
\begin{gathered}
\dot z_n=n[i\omega z_n+h z_{n-1}-h^*z_{n+1}-\gamma (z_n+\tfrac{1}{2}z_{n-1}+\tfrac{1}{2}z_{n+1})]\;.
\end{gathered}
\label{eq:zcnoise}
\end{equation}

\subsection{Effect of Cauchy inhomogeneity}
We further generalize the system under consideration by allowing for non-identical neurons. We assume
that the driving current $I(t)$ possesses an additive quenched quantity $\eta_k$ so that instead of Eq.~\eqref{eq:thcn}, we now consider the TN dynamics
\new{\begin{subequations}
\begin{gather}
\dot\theta_k=1-\cos\theta_k-g\sin\theta_k+(1+\cos\theta_k)\big[I(t)+\Gamma I_k(t)\big] , \label{eq:thcn_inh}
\end{gather}
which, in the corresponding QIF formulation, reads 
\begin{equation}
\dot v_k= v^2_k -gv_k + I(t)+\Gamma I_k(t)\;,
\label{eq:QIFcn_inh}
\end{equation}
with the same resetting as above. 
The neuron-specific input $I_k$ is given by
\begin{equation}
I_k(t) = c \;\! \xi_k(t) + (1-c) \eta_k,\label{eq:individual_input}
\end{equation}
\label{eq:micro_cn_inh}
\end{subequations}
where the parameter $c\in[0,1]$ weights the relative contributions from deterministic heterogeneity and independent noise;
one retrieves Eq.~\eqref{eq:thcn} by setting $c=1$.
We draw the quantities $\eta_k$ (which are similar to natural frequencies
in the Kuramoto model) from a Cauchy-Lorentz distribution $f(\eta)$ with zero mean and unit half-width,
\[
f(\eta)=\frac{1}{\pi(1+\eta^2)}\;.
\]
The parameter $\Delta:= (1-c)\Gamma$ in Eq.~\eqref{eq:micro_cn_inh} determines the spread of ``natural frequencies'', i.e.\ the degree of heterogeneity among neurons.
According to the definition above, the noise intensity is now $\gamma=c\;\!\Gamma$.}

Because the quantities $\eta_k$ are quenched (time-independent), we can consider them as an additional
parameter in the distribution of the variables $\theta$, and write $P(\theta,t;\eta)$ instead of $P(\theta,t)$.
Correspondingly, one can introduce the
order parameters $z_n(t;\eta)$ that now depend also on the values of $\eta$. 
For globally coupled neurons, the mean fields that are relevant then appear after averaging the order parameters over the distribution of $\eta$:
\begin{align*}
Z_n&=\int_{-\infty}^{\infty}d\eta\;f(\eta)z_n(t;\eta)\\
&=\int_{-\infty}^{\infty}d\eta\,f(\eta)\int_0^{2\pi}d\theta\,
P(\theta,t;\eta) e^{in\theta}\;.
\end{align*}

Next we follow the approach of Ott and Antonsen \cite{Ott-Antonsen-08} and make an assumption that the density
$P(\theta,t;\eta)$ is an analytic function of the complex-valued parameter $\eta$ in the upper half-plane and converges exponentially to zero as $\text{Im}(\eta) \to \infty$. 
Then, the integral over $\eta$ can be taken by virtue of the Cauchy residue theorem (using the pole $\eta=i$)
\[
\int_{-\infty}^{\infty}d\eta\,f(\eta)P(\theta,t;\eta)=P(\theta,t;i)\;,
\]
so that
\[
Z_n(t)=z_n(t;i)\;.
\]
Using Eqs.~\eqref{eq:zcnoise} together with $\omega=1+I(t)+\Delta \eta$ and $h(t)=\frac{g+i(I(t)+\Delta\eta-1)}{2}$ 
evaluated at the pole $\eta=i$, we finally obtain
\begin{equation}
\begin{gathered}
\dot Z_n= n\Big[ i(1+I(t))Z_n-(\Delta+\gamma)Z_n+\hspace{1.2cm}\\
+\frac{g+i(I(t)-1)-(\Delta+\gamma)}{2}Z_{n-1}-\hspace{.5cm}\\
-\frac{g-i(I(t)-1)+(\Delta+\gamma)}{2}Z_{n+1}\Big],\quad n\geq 1\;.
\end{gathered}
\label{eq:ni}
\end{equation}

Noteworthy, both Cauchy noise and Cauchy inhomogeneity have the same effect on the dynamics
of the population, \new{that is why we introduced the effective inhomogeneity parameter $\g=\Delta+\gamma$ in Eq.~\eqref{eq:micro_cn_inh}.
We remark, however, that the microscopic state of the network can depend on the relative weighting $c$ of deterministic
and stochastic individual inputs, as analyzed in more detail in \cite{clusella_montbrio_2022}.
We also stress here that} while the effect of noise in Eqs.~\eqref{eq:ni} is unconditional, the effect
of the inhomogeneity is based on an additional assumption about analyticity of the distribution density.
This will be important for the interpretation of the dynamical regimes below.

In another remark we would like to stress a special feature of the Cauchy-distributed noise: because
in Fourier space it acts proportionally to the mode number, the resulting system \eqref{eq:ni} is homogeneous in mode
number $n$, which is essential for the theory below. This was first recognized in Refs.~\cite{Toenjes-Pikovsky-20,Tanaka-20}.
In contradistinction, Gaussian noise acts proportionally to the square of the mode number and thus destroys homogeneity of system \eqref{eq:ni}.

\section{Mean field observables and coupling terms}
\label{sec:obs}
In the description of populations of neurons, the two main observables are typically the mean firing rate $R$
and the mean voltage $V$. In networks of globally coupled neurons, the synaptic input $I(t)$ depends on the 
recurrent coupling and in some (but not all) cases becomes a function of $R$ and $V$. Here, we relate 
the mean firing rate, mean voltage and synaptic input to the order parameters. 

\subsection{Firing rate and mean voltage}
Computational models of large networks of recurrently coupled spiking neurons typically focus on a macroscopic observable that measures the mean rate at which neurons emit spikes, the network firing rate
\begin{equation}
R(t) = \frac1N \sum_{k=1}^N \sum_j \frac1{\tau_r} \int_{t-\tau_r}^t ds \; \delta (s-t_k^{(j)}) \;,
\label{eq:R_fin}
\end{equation}
where the instant $t_k^{(j)}$ corresponds to the $j$-th spike of neuron $k$ (it happens when the corresponding 
variable $\theta_k$ crosses $\pi$), and $\tau_r$ is a time window of spike events. 
Taking first the limit of infinitely many neurons, $N\to \infty$, and then $\tau_r\to0$, one obtains the mean firing rate that, in terms of the probability density in Eq.~\eqref{eq:gfpth}, is defined as the flux of probability density at $\theta=\pi$ \new{(in the Stratonovich
interpretation, the flux at $\theta=\pi$ is purely deterministic, because
the noisy term is multiplied by $(1+\cos\theta)$)}:
\begin{equation}
R(t)=2P(\pi,t)=\frac1\pi \Big[ 1+\sum_{n=1}^\infty (-1)^n(Z_n+Z_n^*)\Big]\;.
\label{eq:frz}
\end{equation}
The mean voltage $V$ is the population average of the membrane potential variables $v$:
\begin{equation}
V(t)=\langle v\rangle=\left\langle \frac{\sin\theta}{1+\cos\theta}\right\rangle=i\sum_{n=1}^\infty (-1)^n(Z_n-Z_n^*)\;;
\label{eq:vz}
\end{equation}
\new{the last equality can be obtained by taking the limit $\lim_{\epsilon\to0} \left\langle \frac{\sin\theta}{1+\cos\theta+\epsilon} \right\rangle$ as in \cite{pietras_etal_19}.}
A combination of both $R$ and $V$ can be expressed as a simple alternating sum of the moments: 
\begin{equation}
\pi R -i V = \new{1+2}\sum\limits_{n=1}^\infty (-1)^n Z_n\;. 
\label{eq:rvz}
\end{equation}
\new{The mapping~\eqref{eq:rvz} between the macroscopic variables $R$ and $V$ and the Kuramoto-Daido order parameters $Z_n$ was originally derived in Eq.~(B2) in \cite{montbrio_pazo_roxin_2015} and could have been used alternatively to obtain the expressions \eqref{eq:frz} and \eqref{eq:vz}.}

\subsection{Recurrent coupling}
Next, we introduce two types of \new{global} recurrent coupling in the population \cite{Ermentrout_06,laing_2015}: electrical coupling via gap junctions and chemical coupling via excitatory/inhibitory synapses.
\new{In the microscopic dynamics Eq.~\eqref{eq:micro_cn_inh}, the input $I(t)$ now incorporates a common external input $I_0$ and two additional terms,
\[
I(t) = I_0 + \frac{g}{N}\sum_{k=1}^N v_k + \frac{\kappa}{N}\sum_{k=1}^N s_k = I_0 + gV - \kappa S\; .
\]
Thus, g}ap junction coupling of strength $g\geq 0$ amounts to including the term $gV$ to the driving
current $I(t)$ of each neuron \cite{pietras_etal_19}.
\new{We model chemical interactions with the term $\kappa s_k$, with $\kappa >0$ $(\kappa<0)$ denoting excitatory (inhibitory) synaptic coupling and the synaptic variables $s_k$ satisfy $\tau_s \dot s_k = \rho(\theta_k) - s_k$, where $\rho(\theta)$ is the spike pulse profile and $\tau_s$ a synaptic time constant. 
The mean synaptic activity $S(t) = \langle s_k \rangle$ satisfies the relaxation equation}
\[
\tau_s \dot S=\langle\rho\rangle -S\;.
\] 
In the limit of fast relaxation, i.e.\ for instantaneous interactions $\tau_s\to 0$, one has $S=\langle \rho \rangle = \int_0^{2\pi} d\theta \rho(\theta)P(\theta,t)$. 

In the following, we briefly present several possible choices of the pulse profile $\rho(\theta)$. We assume that the pulse is localized around $\theta=\pi$ (i.e.~when the membrane potential diverges, $v\to \infty$), 
and that the total area is normalized $\int_{0}^{2\pi}d\theta \rho(\theta)=2\pi$. If we \new{write} the pulse \new{profile} as a Fourier series
\[
\rho(\theta)=\sum_{n=-\infty}^\infty c_n e^{in\theta},\quad c_0=1,
\]
then the average synaptic activity is represented via the order parameters as
\[
\langle \rho\rangle=1+\sum_{n=1}^\infty (c_n^* Z_n + c_n Z_n^*)\;.
\]
\paragraph{Dirac delta-pulses. } Because spikes are rather narrow, in many situations a Dirac $\delta$-function $\rho_{\delta}(\theta)=2\pi\delta(\theta-\pi)$ is adequate. 
The Fourier coefficients are $c_n=(-1)^{|n|}$
and the synaptic activity reduces to the mean firing rate (see Eq.~\eqref{eq:frz})
\begin{equation}
\langle \rho_{\delta}\rangle=1+\sum_{n=1}^\infty (-1)^n(Z_n+Z_n^*)=\pi R(t)\;.
\label{eq:dd}
\end{equation}
\paragraph{Ariaratnam-Strogatz (AS) pulse. } Ariaratnam and Strogatz \cite{ariaratnam2001phase}
suggested a family
of pulse profiles $\rho_{AS}(\theta)=a_m (1-\cos\theta)^m$ with $a_m=\frac{2^m(m!)^2}{(2m)!}$,
where $m$ is an integer parameter. 
The AS pulses are smooth, but in the limit $m\to\infty$ they converge to $\delta$-pulses. The average synaptic activity is a finite sum of the order parameters
\begin{equation}
\langle \rho_{AS}\rangle=1+(m!)^2 \sum_{n=1}^m (-1)^n \frac{ Z_n + Z^\ast_n }{(m+n)!(m-n)!} \;.
\label{eq:as}
\end{equation}
\paragraph{Rectified Poisson (RP) pulse. } Gallego et al.~\cite{Gallego_etal-17} suggested the following
pulse form
\[
\rho_{RP}(\theta)=  \frac{(1-r)(1-\cos\theta)}{1+2r\cos\theta+r^2} \;,
\]
where the ``sharpness'' parameter $ r\in (-1,1)$ determines the width of the pulse:
For $r=-1$, $\rho_{RP}=1$ is flat; for $r=0$, $\rho_{RP}(\theta)=1-\cos\theta$ coincides with the AS pulse with $m=1$; and in the limit $r\to 1$, $\rho_{RP}(\theta)$ becomes a $\delta$-pulse.
The Fourier coefficients are $c_n=\frac{1+r}{2r}(-r)^{|n|}$ and the average activity is the infinite series
\begin{equation}
    \langle \rho_{RP} \rangle = 1 + \frac{1+r}{2r}\sum_{n=1}^\infty (-r)^n \big( Z_n + Z^\ast_n \big) \;.
    \label{eq:rp}
\end{equation}

The different pulse shapes above are symmetric about $\theta=\pi$ and satisfy $\rho(0)=0$.
We remark that the RP pulses of width $r$ can be further generalized to account for non-symmetric pulses when considering Fourier coefficients $c_n=c_{-n}^*$ of the form $c_n=ae^{i(\varphi+n\psi)}r^n, n>0,$ with additional parameters $a>0$ and $\varphi,\psi\in [0,2\pi)$.  
In all these cases, it is possible to represent the relevant observables and mean fields
governing the dynamics of the neural population via the order parameters $Z_n$.
Such representations can directly be used for numerical simulations of the ensemble of neurons in the thermodynamic limit, with a 
proper truncation of the infinite series.

\section{Finite-dimensional reduction}
\label{sec:fdr}
\subsection{Reduction in terms of Kuramoto order parameters}
\label{sec:red}
In this section we \new{demonstrate} that the infinite system of order parameter dynamics \eqref{eq:ni} can be reduced to three complex equations (and a constant function, which is an integral of motion).
Our derivation here directly follows the corresponding derivation for the Kuramoto
problem in \cite{cestnik_pikovsky_2022a}, so we omit some technical details.

First, we rewrite Eqs.~\eqref{eq:ni} in a more compact form
\begin{equation}
\dot Z_n=n\big[(i\omega-\g) Z_n+(h-\tfrac{\g}{2})Z_{n-1}
-(h^*+\tfrac{\g}{2})Z_{n+1}\big],\; n\geq 1.
\label{eq:ni1}
\end{equation}
We introduce the complex-valued exponential generating function (EGF)  $\mathbf{Z}(k,t)=\sum_{n=0}^\infty Z_n(t)\frac{k^n}{n!}$, which obeys the
partial differential equation (a prime denotes derivative with respect to $k$):
\[
\dot{\mathbf{Z}}=k(i\omega-\g)\mathbf{Z}'+k(h-\tfrac{\g}{2})\mathbf{Z}-k(h^*+\tfrac{\g}{2})\mathbf{Z}''\;.
\]
With the Ansatz $\mathbf{Z}(k,t)=e^{kQ(t)}\mathbf{B}(k,t)$ assuming that $Q$ obeys $\dot Q=i\omega Q+h-h^*Q^2 -\frac{\g}{2}(1+Q)^2$,
we find that the generating function $\mathbf{B}(k,t)=\sum_{n=0}^\infty \beta_n(t)\frac{k^n}{n!}$ obeys the dynamics
\[
\dot{\mathbf{B}}=k\left[ i \omega -\g -2(h^*+\tfrac{\g}{2})Q\right] \mathbf{B}'-k(h^*+\tfrac{\g}{2})\mathbf{B}''
\]
and the equations for the new variables $\beta_n$ read
\begin{equation}
\tfrac{1}{n}\dot\beta_n=\left[i\omega-\g -2(h^*+\tfrac{\g}{2})Q\right]\beta_n-(h^*+\tfrac{\g}{2})\beta_{n+1},\; n\geq 1.
\label{eq:bet}
\end{equation}
Due to normalization of the exponential generating function $\mathbf{Z}(0,t)=1$, we have $\beta_0=1$.

We now introduce two new complex dynamical variables $y$ and $s$. As we will show below, the set of variables $\beta_n$ can fully be represented 
through these variables $y,s$ and the constants of motion to be defined below.
The collective dynamics of the network of spiking neurons is exactly described by the following dynamical equations for $Q,y,s$,
\begin{subequations}
\begin{align}
\dot Q &=  i\omega Q + h - h^\ast Q^2 - \tfrac{\g}{2} (1+Q)^2\;, \label{eq:Q}\\
\dot y &=  \big[ i\omega - 2h^\ast Q\big]y - \g(1+Q)y\;, \label{eq:y}\\
\dot s &=h^\ast y + \tfrac{\g}{2} y\;. \label{eq:s}
\end{align}
\label{eq:Qys}
\end{subequations}
To reveal the connection between $\{\beta_n\}$ and $y,s$, we introduce an additional set of variables $\{\alpha_n\}$ via 
\begin{equation}
\beta_n=y^n\alpha_n. 
\label{eq:bya}
\end{equation}
In combination with Eqs.~\eqref{eq:bet} and \eqref{eq:y}, we find the dynamics of $\alpha_n$:
\begin{equation}
\tfrac{1}{n}\dot\alpha_n=-(h^*+\tfrac{\g}{2})y\alpha_{n+1},\quad n\geq 1\;.
\label{eq:alpha}
\end{equation}
In terms of the corresponding ordinary generating function (OGF) $\mathcal{A}(k,t)=\sum_{n=1}^\infty \alpha_n k^n$, 
the dynamics \eqref{eq:alpha} can be represented as
\begin{equation}
\dot{\mathcal{A}}=-(h^*+\tfrac{\g}{2})y\left[\mathcal{A}'-k^{-1}\mathcal{A}\right]\;.
\label{eq:ogfa}
\end{equation}
Finally, we introduce another OGF that is related to $\mathcal{A}$ via the dynamical variable $s(t)$:
\[
\mathcal{M}(k)= \frac{k}{k+s(t)} \mathcal{A}\big(k+s(t),t\big)\;.
\]
A direct calculation of the time derivative of $\mathcal{M}$ yields $\dot{\mathcal{M}}(k)=0$, which means that this 
function is an integral of motion. Equivalently, using the representation $\mathcal{M}(k)=\sum_{n=1}^\infty \mu_n k^n$, we
observe that the system possesses an infinite set of integrals $\mu_n$. To conclude the derivation,
we present the relations between the different variables (see  \cite{cestnik_pikovsky_2022a} for an explicit derivation):
\begin{subequations}
\begin{align}
\mu_n &= \sum\limits_{m=n}^\infty \binom{m-1}{n-1} \alpha_m(t) \big[s(t)\big]^{m-n}\;, \label{eq:mun}\\
\alpha_n &= \sum\limits_{m=n}^\infty \binom{m-1}{n-1} \mu_m \big[-s(t)\big]^{m-n}\;,\label{eq:alphan}\\
\beta_n &= \new{\sum\limits_{m=0}^n \binom{n}{m}Z_m(t)\big[-Q(t)\big]^{n-m}\;,} \label{eq:betaZ}\\
Z_n &=\sum\limits_{m=0}^n \binom{n}{m} \beta_m(t) \big[Q(t)\big]^{n-m} = \label{eq:zn}\\
&=Q^n - \sum\limits_{m=1}^n \binom{n}{m} Q^{n-m} y^m \sum\limits_{d=0}^{m-1} \frac{s^{d-m}}{d!} \mathcal{M}^{(d)} (-s)\;,\nonumber
\end{align}
\label{eq:mualphazn}
\end{subequations}
where $\mathcal{M}^{(d)}$ denotes the $d^\text{th}$ derivative of $\mathcal{M}$ with respect to $k$. 
The Kuramoto order parameter, i.e.~the first circular moment, expresses as:
\begin{equation}
Z_1 = Q-y \frac{\mathcal{M}(-s)}{s} \;.
\label{eq:first_moment}
\end{equation}

\subsection{Reduction in terms of voltage and firing rate}
As we discussed in Section \ref{sec:obs}, macroscopic observables of interest for populations of neurons are the mean firing rate $R$
and the mean voltage $V$.
It is instructive to represent them in terms of the newly introduced dynamical variables $Q,y,s$.
Inserting Eq.~\eqref{eq:zn} into the expansions \eqref{eq:frz} and \eqref{eq:vz} for $R$ and $V$, respectively, and using the definitions
of the OGF $\mathcal{A,M}$, after straightforward algebra we obtain
\begin{subequations}
\begin{align}
R =& \frac{1}{\pi} \text{Re} \Big[ \frac{1-Q}{1+Q} +\frac{2y}{y(1+Q)+s(1+Q)^2}\times \nonumber\\
&\mathcal{M} \left(-\frac{y+s(1+Q)}{1+Q}\right)\Big]\;,\\
V =& -\text{Im}\Big[ \frac{1-Q}{1+Q} +\frac{2y}{y(1+Q)+s(1+Q)^2}\times \nonumber\\
&\mathcal{M}\left(-\frac{y+s(1+Q)}{1+Q}\right)\Big]\;.
\end{align}
\label{eq:rv}
\end{subequations}
Expressions \eqref{eq:rv} suggest a transformation from the variables $Q,y,s$ to new dynamical variables $\Phi,\lambda,\sigma$
according to
\begin{equation}
\Phi=\frac{1-Q}{1+Q},\quad \lambda=\frac{2y}{(1+Q)^2},\quad \sigma=\frac{y}{1+Q}+s\;.
\label{eq:phisl}
\end{equation}
The inverse transformation reads
\begin{equation}
Q=\frac{1-\Phi}{1+\Phi},\quad y=\frac{2\lambda}{(1+\Phi)^2},\quad s=\sigma-\frac{\lambda}{1+\Phi}\;.
\label{eq:phislb}
\end{equation}

When substituting Eq.~\eqref{eq:w_h} for $\omega,h$, we obtain from \eqref{eq:Qys} the dynamical equations for the new variables
\begin{subequations}
\begin{align}
\dot{\Phi} &= i\Phi^2 - g \Phi -i I(t) + \Gamma \;,\label{eq:ndv_phi}\\
\dot{\lambda} &= 2i\Phi\lambda-g\lambda\;, \label{eq:ndv_lambda}\\
\dot{\sigma} &= i\lambda \;.\label{eq:ndv_sigma}
\end{align}
\label{eq:ndv}
\end{subequations}
The main advantage is that the mean firing rate and the mean voltage in terms of these new variables simplify as
\begin{equation}
\pi R - i V = \Phi+\lambda\frac{\mathcal{M}(-\sigma)}{\sigma} \;.
\label{eq:RV_vars}
\end{equation}
\new{
Eq.~\eqref{eq:RV_vars} readily allows for extracting the firing rate or the mean voltage from the dynamical system \eqref{eq:ndv} by taking either the real or the imaginary part of the right hand-side
\begin{subequations}
\begin{align}
R(t) &= \frac1\pi \mathrm{Re} \left[ \Phi + \lambda \frac{\mathcal M(-\sigma)}\sigma \right],\label{eq:r}\\
V(t) &=  \phantom{\frac1\pi}\mathrm{Im} \left[ \Phi + \lambda \frac{\mathcal M(-\sigma)}\sigma \right].\label{eq:v}
\end{align}
\end{subequations}
}

At this point some comments are due:
First, note the resemblance between Eqs.~\eqref{eq:RV_vars} and \eqref{eq:first_moment}, which may be due to the transformal mapping between $\Phi$ and $Q$.
Second, there is also a more apparent resemblance of Eq.~\eqref{eq:ndv_phi} to the initial QIF dynamics~\eqref{eq:QIF1}. 
Third, Eqs.~\eqref{eq:ndv} are asymmetrically coupled: the dynamics~\eqref{eq:ndv_phi} for $\Phi$ does not depend on $\lambda$ and $\sigma$ explicitly, but may implicitly depend on them through the input current $I$, e.g., if $I=I(R,V)$ depends on mean-field terms such as the firing rate $R$ or the mean voltage $V$. 
In case that the current $I$ is not a function of the state variables $\lambda$, $\sigma$, then Eqs.~\eqref{eq:ndv} is a skew system and the first equation~\eqref{eq:ndv_phi} acts as a two-dimensional driving to the other two equations. 

\subsection{Recurrent coupling in terms of dynamical variables}
For a self-consistent description of the collective dynamics, we also need to express the recurrent coupling via gap junctions and/or via the previously introduced synaptic pulses
in terms of the newly introduced variables $\Phi,\lambda,\sigma$. 
As the gap junction coupling term is proportional to the mean voltage $V$, we can here use Eq.~\eqref{eq:v}. Likewise, for synaptic interactions via Dirac $\delta$-pulses $\rho_\delta(\theta)$ according
to Eq.~\eqref{eq:dd}, we can  use  Eq.~\eqref{eq:r} because
the average synaptic activity $\langle \rho_\delta \rangle$ reduces to the mean firing rate $R$. 

In general, however, the synaptic coupling depends on the assumed shape of the pulse
and one obtains complex expressions for $\langle \rho \rangle$,
e.g., for the RP pulse \eqref{eq:rp} we have
\begin{widetext}
\begin{align*}
\langle \rho_{RP} \rangle &= \text{Re} \left[ \frac{1-Q}{1+rQ} + \frac{(1+r)y}{ry(1+rQ)+s(1+rQ)^2} \mathcal{M} \left( -\frac{ry+s(1+rQ)}{1+rQ} \right) \right]\\
&=\text{Re} \left[ \frac{\Phi}{u} + \frac{(1+r)\lambda}{2\sigma u^2-(1-r)\lambda u} \mathcal{M}\left( -\sigma+\frac{(1-r)\lambda}{u} \right)\right]\;,
\end{align*}
where we introduced the auxiliary variable $u = \frac{1}{2}(1+r+(1-r)\Phi) = \frac{1+rQ}{1+Q}$ with $u\big|_{r=1}=1$. 

For the AS pulse, $\langle \rho_{AS} \rangle$ is represented by a finite sum of order parameters \eqref{eq:as}. Here one has to use Eq.~\eqref{eq:zn} expressing these order parameters via dynamical variables $Q,y,s$. If needed, the transformation~\eqref{eq:phislb}
to variables $\Phi,\lambda,\sigma$ can be used.
\end{widetext}

\subsection{Initial conditions}\label{sec:init_cond}
Given an initial distribution $P(\theta,t=0)$ of phases $\theta(0)$, or $W(v,t=0)$ of membrane potentials $v(0)$, we ought to initialize the dynamical system $(Q,y,s)$ accordingly. 
Note however, that the set of dynamical variables $Q,y,s$ is underdetermined, \new{which is characteristic for a finite-dimensional theory of infinite-dimensional dynamics. There is a certain freedom in choosing the initial values $Q(0),y(0),s(0)$, which follows from the way we introduced them in relation to the moments $Z_n$. 
For any set $(Q(0),y(0),s(0))$, one can find corresponding values of $\mu_n$. Indeed, a 
straightforward combination of relations~\eqref{eq:mun}, \eqref{eq:bya}, and \eqref{eq:betaZ} begets an explicit formula relating constants $\mu_n$ (which in turn define the constant function $\mathcal{M}(k)$) 
to $Q(0),y(0),s(0)$ and $Z_n(0)$:
\begin{gather*}
\mu_n = \frac{1}{\left[s(0)\right]^n} \sum\limits_{m=n}^\infty \binom{m-1}{n-1} \left[-\frac{s(0)Q(0)}{y(0)}\right]^m\times\\ \sum\limits_{j=0}^m \binom{m}{j} Z_j(0) \big[ -Q(0) \big]^{-j}\;.
\end{gather*}
Of course,
the dynamical evolution of order parameters $Z_n(t)$ is the same for all admissible choices. This arbitrariness has been already discussed by Watanabe and Strogatz \cite{Watanabe-Strogatz-94} in their finite-dimensional reduction for the population of identical oscillators. Here, we} 
adopt the approach of \cite{cestnik_pikovsky_2022a} 
and define the initial values as
\begin{equation}
Q(0)=s(0)=0,\quad y(0)=1.
\label{eq:ic1}
\end{equation}
In this case, as it follows from relations~\eqref{eq:mualphazn}, 
\[
\mu_n=\alpha_n(0)=\beta_n(0)=Z_n(0)\;.
\]
The constant OGF $\mathcal{M}(k)$ is thus directly related to the initial values of the order parameters
\[
\mathcal{M}(k)=\sum_{n=1}^\infty Z_n(0) k^n\;.
\]
Substituting here the expression of the order parameters via the initial distribution density of the phase variables $P(\theta,0)$, we obtain
\begin{equation}
\begin{gathered}
\mathcal{M}(k)=\int\limits_0^{2\pi}d\theta\; P(\theta,0)\frac{ke^{i\theta}}{1-ke^{i\theta}}=\\
=\int\limits_{-\infty}^\infty dv\;W(v,0)\frac{k(1+iv)}{1-k-iv(1+k)}\;.
\end{gathered}
\label{eq:initm}
\end{equation}
In terms of the observables $\Phi,\lambda,\sigma$ the initial conditions are
\begin{equation}
\Phi(0)=\sigma(0)=1,\quad \lambda(0)=2\;.
\label{eq:icnv}
\end{equation}
Below we present several cases where an expression for $\mathcal{M}(k)$ is relatively simple.
\begin{itemize}
\item A delta-distribution of voltages $W(v,0) = \delta(v - v_0)$ corresponds to $P(\theta,0) = \delta(\theta - \theta_0)$ with $\theta_0 = 2 \arctan(v_0)$, leading to $Z_n = e^{in\theta_0}$ and thus $\mathcal M(k) = \frac{e^{i\theta_0} k}{1-e^{i\theta_0} k} = k\big( \frac{1+iv_0}{1-iv_0}-k \big)^{-1}.$
\item A Cauchy-Lorentz distribution of voltages 
\begin{equation}
W(v,0) = \frac{1}\pi \frac{x}{(v-v_0)^2 + x^2} \text{ for some }x \in \mathbb R_{>0}
\label{eq:cdv}
\end{equation}
 corresponds to a wrapped Cauchy distribution of the phase variables 
$P(\theta,0) = \frac1{2\pi} \frac{1-|\mu|^2}{|1-\mu e^{-i\theta}|^2}$ where $\mu = \frac{1 - \pi x + i v_0}{1 + \pi x - i v_0} \in \mathbb C$.
This leads to $\mathcal M(k) = \frac{\mu k}{1-\mu k}$. The initial order parameters are powers of the parameter, $Z_n(0)=\mu_n = \mu^n$. 

A skewed generalization of the Cauchy distribution can also easily be represented with the Kato-Jones~\cite{kato-jones} distribution of phases, 
which yields $\mathcal{M}(k) = c \frac{\mu k}{1-\mu k}$ where $c \in \mathbb C$. 
\item A uniform distribution of voltages in some interval $[v_0-d,v_0+d]$ with $v_0 \in \mathbb R, d>0$,
\begin{equation}
W(v,0) = \begin{cases} \frac{1}{2d}\quad &\text{if } v \in [v_0-d,v_0+d],\\
0, &\text{otherwise},
\end{cases}
\label{eq:udv}
\end{equation}
corresponds to $P(\theta,0)=1/(2d)/(1+\cos\theta)$ if $-\pi < 2 \arctan(v_0-d) \leq \theta \leq 2\arctan(v_0+d) < \pi$ and $0$ otherwise,
where we used that $\arctan(v)$ is bijective on $(-\pi,\pi)$. 
The OGF is
\begin{equation}
\begin{aligned}
&\mathcal M(k) = \frac{-1}{d} \frac{k}{(1+k)^2} \Big[ (1+k)d + \\
&\hspace{0.4cm}+ i \log\left( \frac{(1+k)(v_0-d+i)-2ik}{(1+k)(v_0+d+i)-2ik}\right) \Big] .
\end{aligned}
\label{eq:ogfun}
\end{equation} 
\end{itemize}
We notice also that a weighted sum of different initial distributions
allows for a proper representation of the constant OGF $\mathcal{M}(k)$. 
Because
of linearity of representation~\eqref{eq:initm}, if the initial density of the phase
variable is a sum of ``elementary'' densities, then the
OGF is a sum of ``elementary'' OGFs: 
\begin{equation}
P(\theta,0)=\sum_m w_m P_m(\theta,0),\quad \mathcal{M}(k)=\sum_m w_m \mathcal{M}_m(k)
\label{eq:mix}
\end{equation}
for weights $w_m\geq0$ with $\sum_m w_m=1$.
We will explore 
this property in the next Section \ref{sec:identical} and in the Section \ref{sec:nex} below, where we discuss the dynamics of a population
of neurons following the resetting of a fraction of neurons to a new state.

Finally, we mention that the function $\mathcal{M}(k)$ can be approximated for an arbitrary distribution $W(v,0)$, by first expressing it in the TN formulation $P(\theta,0)$ via~\eqref{eq:ptow}, and then approximating it with a finite Fourier series: $P(\theta) \approx \frac{1}{2\pi} \big[ 1+ \sum\limits_{n=1}^N a_n e^{-in\theta} + \text{c.c.} \big]$, yielding a polynomial $\mathcal{M}(k) \approx \sum\limits_{n=1}^N a_n k^n$.

\subsection{Identical units without noise}\label{sec:identical}
In the case of identical units without noise, $\g = 0$, the following simplification follows from Eq.~(35) in \cite{cestnik_pikovsky_2022a}:
\begin{equation}
\lambda = (\Phi+\Phi^*)\sigma\;.
\end{equation}
The dynamics~\eqref{eq:ndv_lambda} for $\lambda$ thus becomes redundant and Eq.~\eqref{eq:ndv_sigma} for $\sigma$ reduces to $\dot{\sigma} = i (\Phi+\Phi^*) \sigma = 2i \text{Re}[\Phi]\sigma$. 
Given the initial value $\sigma(0)=1$, we thus have $\sigma(t)=\exp[i\zeta(t)]$ and the argument $\zeta\equiv \arg(\sigma)$ follows the real-valued dynamics $\dot{\zeta} = 2\text{Re}[\Phi]$.
Hence, for $\g = 0$ the collective dynamics is 3-dimensional. 
Note that the relation~\eqref{eq:RV_vars} connecting the firing rate and mean voltage remains valid as well as the different choices of initial conditions described in Section~\ref{sec:init_cond}. 
The full three-dimensional dynamics can be summarized as:
\begin{subequations}
\begin{align}
\dot{\Phi} &= i\Phi^2-g\Phi-iI(t)\;,\\
\dot{\zeta} &= 2\text{Re}[\Phi]\;,
\end{align}
\label{eq:ndv_ws}
\end{subequations}
with initial conditions $\Phi(0) = 1$, $\zeta(0) = 0$; and the firing rate $R$ and mean voltage $V$ can be expressed via Eq.~\eqref{eq:RV_vars} as:
\begin{equation}
\pi R -i V = \Phi + (\Phi+\Phi^*)\mathcal{M}(-e^{i\zeta})\;.
\end{equation} 
This essentially boils down to the Watanabe-Strogatz (WS) description~\cite{watanabe_strogatz_1993,Watanabe-Strogatz-94}
of a population of identical units, here translated for QIF neurons. The constant function $\mathcal{M}(k)$ contains the information of the WS constants of motion $\psi_j = \theta_j(0)$, which are, for this simple choice of initial conditions~\eqref{eq:icnv}, just the initial values of the TN phases and relate to the initial voltages via $v_j(0) = \tan(\psi_j/2)$. %
The function $\mathcal{M}(k)$ is in general expressed as: 
\[
\mathcal{M}(k) = \sum\limits_{n=1}^\infty \av{e^{in\psi_j}} k^n = \sum\limits_{n=1}^\infty \av{\left( \frac{1+iv_j(0)}{1-iv_j(0)} \right)^n} k^n \;.
\] 

This description outlined above is also applicable for finite networks: 
Consider an ensemble of $N$ identical neurons $\{\theta_1,\theta_2,..., \theta_N\}$ (or $\{v_1,v_2,...,v_N\}$ in the QIF representation), whose distribution density can be described as a sum of delta functions: $P(\theta) = \frac{1}{N}\sum\limits_{n=1}^N \delta(\theta-\theta_n)$ (or $W(v) = \frac{1}{N}\sum\limits_{n=1}^N \delta(v-v_n)$). 
For each unit $\theta_n$ (or $v_n$), we can determine the constant function $\mathcal M_n(k)$ corresponding to the respective delta function, that is, $\mathcal{M}_n(k) = \frac{e^{i\theta_n(0)}k}{1-e^{i\theta_n(0)}k} = k\big(\frac{1-iv_n(0)}{1+iv_n(0)}-k\big)^{-1}$, $n=1,\dots,N$. Using Eq.~\eqref{eq:mix}, we thus obtain the full function $\mathcal{M}(k) = \frac{1}{N} \sum\limits_{n=1}^N \mathcal{M}_n(k)$. In this way, ensembles with an arbitrary number of QIF neurons can be integrated exactly in just 3 dimensions as their collective dynamics is restricted to the three-dimensional manifold with coordinates $|\Phi|, \arg(\Phi), \zeta$. We stress again that this three-dimensional description is only valid for identical neurons without noise, $\g = 0$, i.e.~without dissipation.

\section{Ott-Antonsen manifold and its stability}
\label{sec:oa}
\subsection{Ott-Antonsen and Lorentzian manifolds}
From Eqs.~\eqref{eq:Qys} it is obvious that $y=0$ is a solution. 
The manifold $\{y=0\}$ is invariant, and on it
the variable $s$ becomes irrelevant because, as it follows from Eq.~\eqref{eq:zn}, the order parameters
are just powers of the variable $Q$: $Z_n=Q^n$. This manifold is thus completely described by the complex variable $Q$, and hence is two-dimensional. 
It has first been identified by Ott and Antonsen~\cite{Ott-Antonsen-08}, who subsequently demonstrated that, in the case of heterogeneous oscillators $\omega_j\neq\omega_k$ with common forcing $h_j=h$, it is attractive in the time-asymptotic limit~\cite{Ott-Antonsen-09,ott2010comment}.
Due to the relation $Z_n=Q^n$, the distribution $P(v,t)$ of phases $\theta$ on the so-called OA manifold follows a wrapped Cauchy distribution (a.k.a.\ Poisson kernel).

In terms of the variables $\Phi,\lambda,\sigma$, the OA manifold corresponds to $\{\lambda=0\}$ and on it the dynamics are fully captured by the variable $\Phi = \pi R - i V$. \new{For $\lambda=0$, the voltages $v$ follow a Cauchy-Lorentz distribution, so} in this context one calls the OA manifold also the Lorentzian manifold~\cite{montbrio_pazo_roxin_2015}.
The dynamics of neural populations on the OA, or Lorentzian, manifold has been explored
in a variety of setups, building primarily on the original results in Refs.~\cite{Luke-Barreto-So-13,So-Luke-Barreto-14,Laing-14,montbrio_pazo_roxin_2015,laing_2015}. 
Time-asymptotic attractiveness of 
the Lorentzian manifold was later proven in \cite{pietras_daffertshofer_2016}.

\subsection{Global stability of the reduced OA manifold}
Here we \new{demonstrate} that the reduced OA manifold for TN, or the Lorentzian manifold for QIF neurons, 
is asymptotically stable for $\g>0$, i.e.~in presence of an inhomogeneity of input currents $\eta$ or of Cauchy white noise $\xi$.
In fact, we show that asymptotically $y\to 0$ so that the OA manifold is attractive. Our starting point are Eqs.~\eqref{eq:Q},\eqref{eq:y}.
We introduce two real variables $X=1-|Q|^2$ and $Y=|y|^2$, and express their dynamics as
\begin{subequations}
\begin{align}
\frac{\dot{X}}{X} &= -(h^*Q+h Q^*) +2\g \frac{1-X}{X}+\g(Q+Q^*)\frac{2-X}{2X}\;, \\
\frac{\dot{Y}}{Y} &= -(h^*Q+hQ^*) -\g (1+\frac{Q+Q^*}{2})\;.
\end{align}
\label{eq:XY}
\end{subequations}
First, we argue that $X$ cannot become negative. Indeed, because at $X=0$, $|Q|=1$, $2+Q+Q^*\geq 0$ and therefore $X$ cannot become negative. Moreover, $X$ cannot vanish because for this one needs the state with $Q=-1$ to be steady, but at this value of $Q$ from Eq.~\eqref{eq:Q} we have $\dot Q\big|_{Q=-1}=-2i$ independently
of the coupling terms $g,I$. This corresponds to the fact that in the original formulation \eqref{eq:TN1} at $\theta=\pi$ one has $\dot\theta=2$ independently of the forcing terms $g,I$.

Next, we combine these two equations in one:
\begin{equation}
\frac{\dot{Y}}{Y} =\frac{\dot{X}}{X} - \g \frac{|1+Q|^2}{1-|Q|^2}\;.
\end{equation}
We integrate this equation to obtain
\begin{equation}
Y(t)=Y(0)\frac{X(t)}{X(0)}\exp[-\g \int\limits_0^t \frac{|1+Q(t')|^2}{1-|Q(t')|^2}\,dt' ] \;.
\label{eq:yxint}
\end{equation}
If $|Q(t)+1|$ is bounded from zero, then also $\frac{|1+Q(t)|^2}{1-|Q(t)|^2}$ has a lower positive bound,
and so the integral in Eq.~\eqref{eq:yxint} tends to zero exponentially for $\g>0$. On the other hand, if $Q\to -1$, then $X$ vanishes
and in this case $Y$ vanishes as well. Thus in all the cases $Y(t)$ eventually vanishes, which means convergence of arbitrary
initial states to the OA manifold.
Furthermore, $Y\to 0$ entails $y\to 0$, which in turn entails $\lambda\to0$ due to Eq.~\eqref{eq:phisl}, thus proving global stability also of the Lorentzian manifold described by $\Phi$.

The convergence $Y\to 0$ is ensured for $\Gamma>0$, but for vanishing noise  and/or inhomogeneity $\Gamma=0$ we obtain $Y(t)=X(t)\frac{Y(0)}{X(0)}$. Thus, $Y$ can vanish only if eventually $X\to 0$, which corresponds to a delta-function distribution with $|Q|=1$, i.e.~to full synchrony of the neurons. In all other situations the OA manifold is not attracting.

\section{Dynamics off the OA manifold}
\label{sec:nex}
As we have discussed in the previous Section \ref{sec:oa}, asymptotically the six-dimensional dynamics reduces
to a two-dimensional dynamics on the OA manifold. Thus, the exact six-dimensional evolution derived above 
is relevant for transient processes only, as the attractors themselves lie on the OA manifold.
Importantly, the full six-dimensional dynamics is needed to faithfully determine the basins of attraction of different asymptotic regimes.
The two-dimensional OA \new{theory} is restricted to specific initial conditions that already lie on the invariant OA manifold (wrapped Cauchy distribution for TN phases, Cauchy-Lorentz distribution for QIF voltages).
By contrast, our approach captures the exact six-dimensional evolution from general initial phase/voltage distributions, which are incorporated in the corresponding $\mathcal M$-function as described in Section~\ref{sec:init_cond}.
In the following, we present three examples that underscore the power of our exact finite-dimensional reduction.

\new{
\begin{figure*}
\centering
\includegraphics[width=0.8\textwidth]{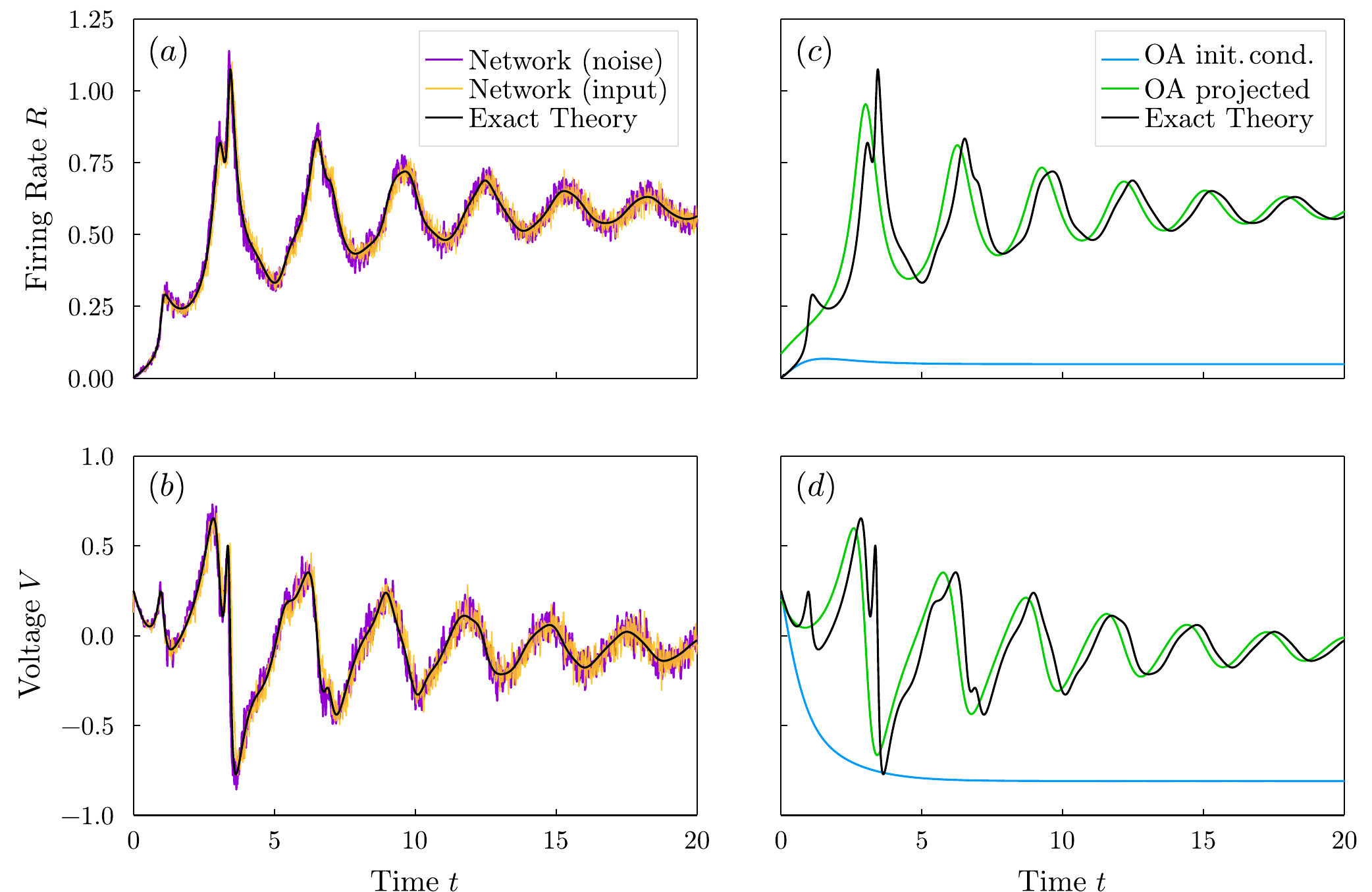}
\caption{\new{Numerical simulations of spiking neuron networks, the exact six-dimensional theory and the two-dimensional OA theory.
Initial voltages $v_j(0)$ are uniformly distributed on $[v_0-d,v_0+d]$ with $d=1, v_0=0.25$. 
(a) Firing rate $R(t)$ and (b) mean voltage $V(t)$ obtained from the microscopic network Eq.~\eqref{eq:micro_cn_inh} with $N=10^5$ neurons driven by Cauchy noise ($c=1$, violet curves) or with Cauchy-Lorentz distributed inputs ($c=0$, yellow curves) are in excellent agreement with the exact low-dimensional theory Eqs.~\eqref{eq:ndv} and \eqref{eq:ogfun} (black curves).
(c,d) Comparison of the exact theory (black, the same curves
as in panels (a,b)) with the two-dimensional OA dynamics initialized with $R(0)=0$ and $V(0)=v_0$ according to the initial voltage distribution (blue) or when projecting the initial voltage distribution on the OA manifold (green), $R(0)$ and $V(0)$ are then obtained from the Kuramoto order parameter $Z_1(0)$ via $\pi R(0) - iV(0) = (1-Z_1(0))/(1+Z_1(0))$.
In panels (a,b) the network firing rate and the 
voltage are smoothed with a rectangular filter of width $0.025$.
Other parameters: noise/heterogeneity strength $\Gamma=1/4$, coupling strengths $\kappa=15\sqrt{\Gamma}/\pi$, input current $I_0=-4\Gamma$ and $g=0$.}
}
\label{fig:sim}
\end{figure*}
}


\subsection{Complex initial transients of the collective dynamics}

As the first litmus test, we compare our finite-dimensional reduction Eqs.~\eqref{eq:ndv} to numerical simulations of large networks of spiking neurons, see Fig.~\ref{fig:sim}.
For the network simulations, we consider \new{$N=10^5$ neurons either driven by Cauchy white noise or by heterogeneous inputs drawn from a Cauchy-Lorentz distribution, whose dynamics are given by Eq.~\eqref{eq:micro_cn_inh} with $c=1$ or $c=0$, respectively.}
The neurons are all-to-all coupled via instantaneous Dirac $\delta$-pulses (Eq.~\eqref{eq:dd}) of strength $\kappa>0$, and $g=0$.
The initial voltages $v_j(0)$ are uniformly distributed in $[v_0-d,v_0+d]$ with $d=1$, so that the initial firing rate $R(0)=0$ and mean voltage $V(0)=v_0$.
The population firing rate $R(t)$ can be computed according to Eq.~\eqref{eq:R_fin}, or in the TN framework as
$R(t) = \frac1{N\; dt} \sum_{j=1}^N 1_{(\pi-2 dt,\pi]}\big(\theta_j(t)\big)$,
where $1_A(\theta)$ is the indicator function and the interval $A=(\pi-2 dt,\pi]$ ensures that we count all neurons that cross the firing threshold $\theta=\pi$ within the next integration step $dt$, and thus elicit a spike before time $t+dt$.
The mean voltage $V(t)$ is computed as the average over the neurons' voltages, or in the TN framework as $V \approx \langle \frac{\sin \theta}{1+\cos\theta +\epsilon}\rangle$ with $\epsilon =10^{-5}$; note that when following~\cite{Ermentrout_06,laing_2015} with $\epsilon=10^{-2}$, there will be some initial disagreement between network voltage and theory.

The collective dynamics of the network is described by Eqs.~\eqref{eq:ndv} with $I(t) = I_0 + \kappa\pi R(t)$, which is exact in the thermodynamic limit with initial conditions $\Phi(0) = \sigma(0) = 1$, $\lambda(0)=2$.
The initial uniform voltage distribution amounts to the $\mathcal M$-function given by Eq.~\eqref{eq:ogfun} with $d=1$.
As shown in Fig.~\ref{fig:sim}(a) and (b), the match between \new{network simulations (violet with Cauchy noise and yellow with heterogeneous inputs) and exact theory (black) is remarkable. 
Even the complex initial transient (up to time $t\approx 8$) is excellently captured by our theory.
By contrast, the two-dimensional OA theory cannot account for such a perfect agreement (Fig.~\ref{fig:sim}c,d).
To begin, it is unclear how to choose the initial conditions on the OA manifold.
In the OA theory, the width parameter $d$ of the initial uniform voltage distribution is lost, so that it is unclear whether the OA dynamics describes the network evolution from the correct initial condition ($d=1$) or even from a delta-distribution of voltages ($d=0$).

The choice of initial conditions, however, can have significant consequences for the predicted collective dynamics---especially in bistable regimes as considered here, see Fig.~\ref{fig:mex}(a) for the corresponding bifurcation diagram.
When initializing the OA dynamics (Eqs.~\eqref{eq:ndv} with $\lambda \equiv0$) according to the initial conditions $R(0)=0$ and $V(0)=v_0$, which corresponds to the initial microscopic network state, then the dynamics predicted by the OA theory can run into an attractor that is different from the actual network dynamics (blue curve in Fig.~\ref{fig:sim}c,d).
We recover the correct attractor of the collective dynamics when projecting the initial voltage distribution onto the OA manifold by, first, computing the Kuramoto order parameter $Z_1(0) = 1/N \sum_{j=0}^N \exp(2i\arctan v_j(0))$ and, second, determining the initial $R$ and $V$ values on the OA manifold via $\pi R(0) - iV(0) = (1-Z_1(0))/(1+Z_1(0))$, see Eq.~\eqref{eq:rv} with $y=0$.
The transient OA dynamics evolves 
towards the true attractor (green curve in Fig.~\ref{fig:sim}c,d); however, it 
does not completely coincide with our exact six-dimensional theory nor with the microscopic network dynamics.
The damped oscillations exhibit a phase lag between OA dynamics and our exact theory.
Moreover, since the OA theory is restricted to two-dimensional dynamics, its behaviour is limited to either an almost monotonous decay (blue curves in Fig.~\ref{fig:sim}c,d) or a simple oscillatory decay with a monotonically decaying amplitude (green curves). By contrast, the full six-dimensional dynamics can exhibit many modes in the transient decay patterns (black curves). 
}

\subsection{Resetting a fraction of neurons induces switching between attractors}
As the attentive reader may have noticed, in the example above a slight change in \new{initial conditions, while keeping all the other parameters the same, resulted} in collective dynamics converging to different attractors:
\new{either an asynchronous low-activity state or an asynchronous high-activity state; synchronous behavior would correspond to collective oscillations.}
As mentioned above, the attractors of the network dynamics lie on the two-dimensional OA manifold, allowing for a concise bifurcation analysis of the bistability between the low- and high-activity states.
Below we expand on this bistable situation, which was already reported in~\cite{montbrio_pazo_roxin_2015} and where the basins of attraction were explored within the OA manifold. Here we show how our approach generalizes the basins of attraction in the full state space of QIF spiking neurons driven by Cauchy white noise and how resetting a fraction of neurons can induce a switch from one to the other attractor.

\begin{figure}[h!]
\centering
\includegraphics[width=0.45\textwidth]{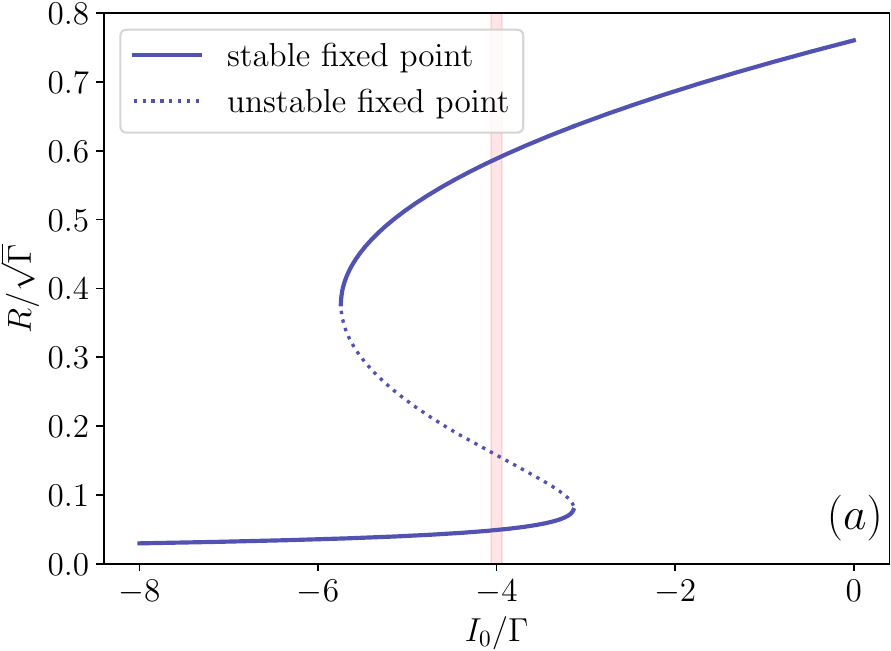}\\[0.2cm]
\includegraphics[width=0.4\textwidth]{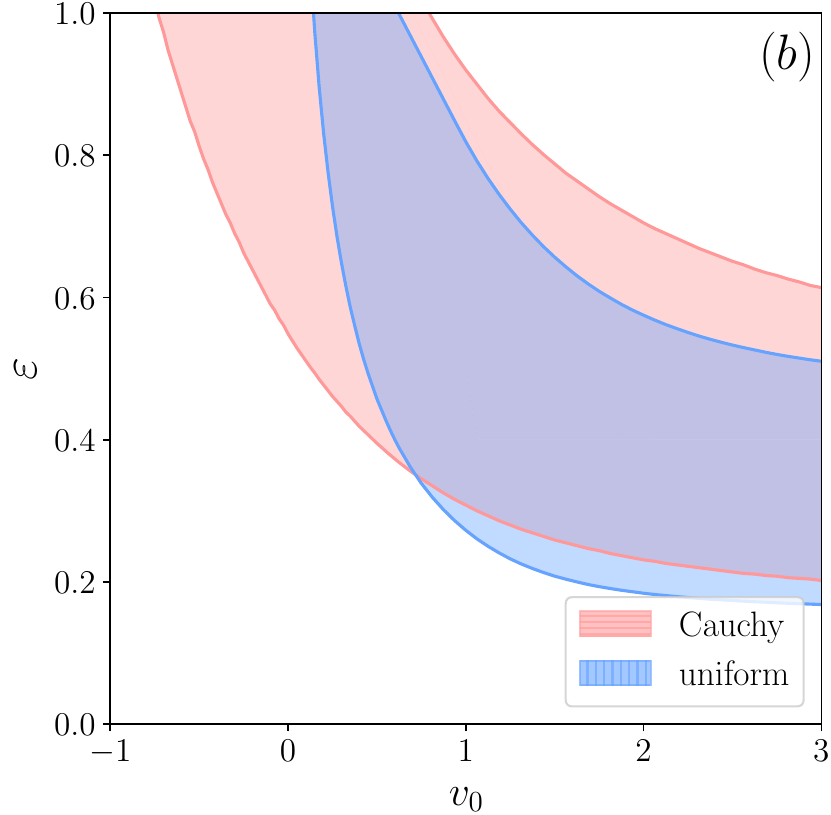}
\caption{Attractor switching due to resetting of a fraction of neurons. (a): Bifurcation analysis yields a region of bistability between two stable, stationary states (full line) whose basins of attraction on the OA manifold are separated by the \new{stable manifolds of the} 
unstable stationary state (dotted line). The pink vertical line marks the chosen parameter regime $I_0/\Gamma = -4$. (b): Basin boundaries of high-activity attractor: after initializing the system in the low-activity state, we reset a fraction $\varepsilon$ of oscillators according to a uniform distribution (blue) or a Cauchy distribution (pink), centered at $v_0$ and having half-width (at half-maximum) 1. Shaded regions indicate switching to the high-activity state. White domains correspond to decay after the perturbation back to the initial low-activity state.
Model parameters as in Fig.~\ref{fig:sim}: $\tau_s = 0$, $g=0$, $\Gamma = 0.25$, $I_0 = -4\Gamma$ and $\kappa = 15\frac{\sqrt{\Gamma}}{\pi}$. Runge-Kutta 4th order method with time step $dt = 10^{-3}$ was used for integration. }
\label{fig:mex}
\end{figure}

As before, we consider global recurrent coupling via instantaneous Dirac $\delta$-pulses and set $g=0$, so that the total input current is $I(t)=I_0+\kappa\pi R(t)$. In dependence on the scaled parameter $I_0/\g$ one observes a bistability of the stationary firing rate,
as shown in Fig.~\ref{fig:mex}(a).
This bistability scenario corresponds to that of Fig.~1(b) in Ref.~\cite{montbrio_pazo_roxin_2015}, where the neurons were assumed heterogeneous ($\Delta=1/4$) without noise ($\gamma=0$). \new{In our notation, their setting corresponds to $\Gamma=1/4$ and $c=0$ in Eq.~\eqref{eq:micro_cn_inh}, but we stress that the same bifurcation diagram can be achieved for any (microscopic) weighting $c\in[0,1]$ between Cauchy heterogeneity and noise. In the following, we set $c=1$ and the coupling strength is $\kappa=\frac{15}{\pi}\sqrt{\g}$.}

We now induce switching from the low- to the high-activity state by resetting a fraction $\varepsilon$ of the neurons to a predefined voltage distribution.
We consider here identical, noise-driven neurons \new{($c=1)$}; this approach cannot be pursued in the presence of heterogeneity \new{($c<1)$} because then the selection of neurons that are reset crucially influences the collective dynamics.
The common input is fixed at $I_0/\g=-4$ in the region of bistability and we initialize the neurons on the low-activity branch.
At the fixed point, the distributions $W_0(v)$ of voltages 
$v$ and $P_0(\theta)$
of phase variables $\theta$ is Cauchy-Lorentz, or wrapped Cauchy, respectively, as it should be on the OA manifold. Then we take a portion $0\leq \varepsilon \leq 1$ of the neurons and redistribute their voltages (phases) according to a new distribution $W_1(v)$ ($P_1(\theta)$, correspondingly). Thus, according to expression \eqref{eq:mix}
we have to start calculations of the transient using full Eqs.~\eqref{eq:ndv} with $\mathcal{M}(k)=(1-\varepsilon)\mathcal{M}_0(k)+\varepsilon\mathcal{M}_1(k)$. (Note that in Ref.~\cite{Gupta-resetting} resettings inside the OA manifold have been considered).

We use two distributions $W_1(v)$ for the resetting:
\begin{enumerate}  
\item A Cauchy-Lorentz distribution \eqref{eq:cdv} with fixed half-width at half-maximum $x=1$ and varying parameter $v_0$.
Then, $\mathcal{M}_1(k)= \frac{\mu k}{1-\mu k}$ with $\mu = \frac{1 - \pi  + i v_0}{1 + \pi  - i v_0}$.
\item A uniform distribution of voltages \eqref{eq:udv} with half-width $d=1$ and different $v_0$.
Because at the initial condition $\sigma(0)=1$, the denominator of $\mathcal{M}_1(-\sigma)$
vanishes, one needs an expansion \begin{equation}\label{eq:M_expanded}
\mathcal{M}_1(-\sigma) = -\frac{1+iv_0}{2} - \frac{3+\delta^2+3v_0^2}{12}(\sigma-1) + \mathcal{O}((\sigma-1)^2)
\end{equation}
to start calculations properly.
\end{enumerate}
Integrating the system of equations \eqref{eq:ndv} with the elaborated constant functions $\mathcal{M}(k)$, we have observed
convergence to one of the attractors on the OA manifold. The basins of attraction for the different resetting distributions $W_1(v)$
are depicted in the parameter plane $(v_0,\varepsilon)$ in Fig.~\ref{fig:mex}(b).
Resetting a large fraction of neurons according to a Cauchy-Lorentz distribution (red), significantly increases the chances of attractor switching from the low- to the high-activity state compared to a resetting according to a uniform distribution (blue).

\subsection{Phase-dependent resetting from synchronous to asynchronous states}
\begin{figure}[h!]
\centering
\includegraphics[width=0.45\textwidth]{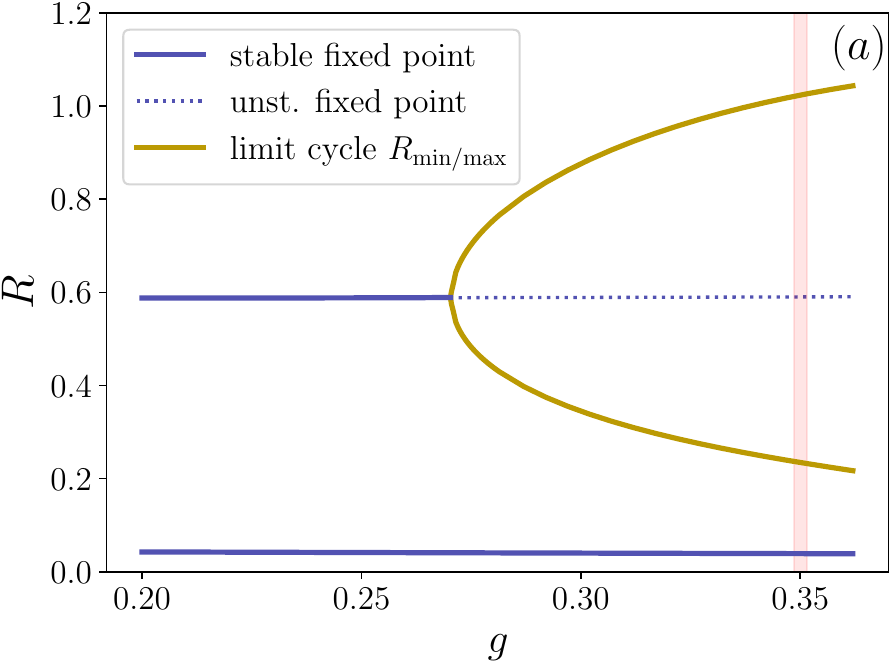}\\[.2cm]
\includegraphics[width=0.4\textwidth]{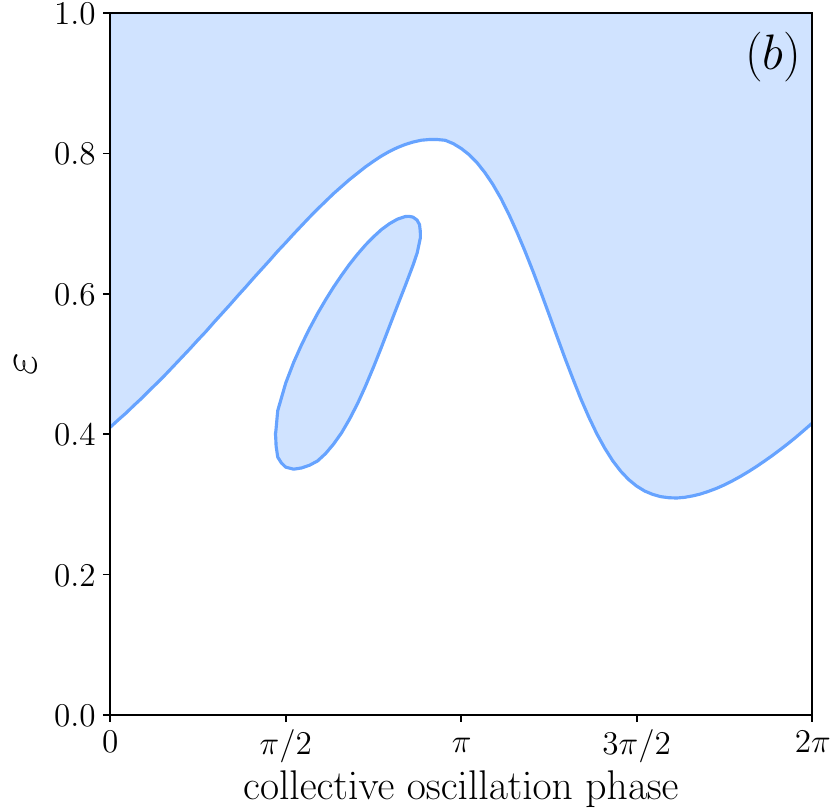}
\caption{Phase-dependent attractor switching from synchronous to asynchronous states. (a): Bifurcation analysis yields a region of bistability between an oscillatory, synchronous state (orange), which emerges from a stable high-activity state via a supercritical Hopf bifurcation, and a stable low-activity stationary state (full line). The pink vertical line marks the chosen parameter regime $g=0.35$. (b): Basin boundaries of low-activity attractor: after initializing the system in synchronous state, we reset a fraction $\varepsilon$ of oscillators according to a uniform distribution centered at $0$ and having half-width 1 for every phase of the collective oscillation. Shaded blue regions correspond to values that induce a switch to the fixed point. White domain corresponds to decay after the perturbation back to the collective oscillation state.
Model parameters are: $\tau_s = 0$, $g=0.35$, $\Gamma = 0.25$, $I_0 = -4\Gamma$ and $\kappa = 15\frac{\sqrt{\Gamma}}{\pi}$. Runge-Kutta 4th order method with time step $dt = 10^{-3}$ was used for integration.}
\label{fig:rot}
\end{figure}

We now add gap junction coupling $g>0$ while keeping the other parameters as before.
For gap junctions, the input current now reads $I(t)=I_0+\kappa\pi R(t) + gV(t)$.
Upon increasing $g$, the high-activity state undergoes a Hopf bifurcation and becomes a stable limit cycle on the OA manifold; the low-activity state remains a phase locked, fixed point solution, see Fig.~\ref{fig:rot}(a). 
We now fix $g=0.35$ and initialize the system in the oscillatory, synchronous steady state, $Q(t)=Q(t+T)$ with period $T$, that has bifurcated from the high-activity branch. 
At different phases of the collective oscillation, 
we reset a portion $\varepsilon$ of oscillators with a uniform distribution of voltages with mean voltage $v_0 = 0$ and half-width $d = 1$. 
We therefore consider the function $\mathcal{M}(k)=(1-\varepsilon)\mathcal{M}_0(k)+\varepsilon\mathcal{M}_1(k)$ as a combination of the Cauchy $\mathcal{M}_0(k) = \frac{Q k}{1-Qk}$ with the expression~\eqref{eq:ogfun} for the $\mathcal{M}_1(k)$. 
In this special case ($v_0 = 0$, $d = 1$), $\mathcal{M}_1(k)$ is given by
\begin{equation}
\mathcal M_1(k) = \frac{-k}{(1+k)^2} \Big[ (1+k) - \pi/2 - 2\arctan(k) \Big] 
\label{eq:M_vuni}
\end{equation}

The basin boundary of the asynchronous low-activity state is depicted in blue in Fig.~\ref{fig:rot}(b), and sensitively depends on the collective oscillation phase and on the reset fraction $\varepsilon$ of neurons.
By contrast, resetting according to a Cauchy-Lorentz distribution $W_1(v)$ centered at 0 with half-width at half-maximum 1 never induces a switch (for any portion $\varepsilon$ and at any phase of collective oscillation). 

\section{Discussion \& Conclusion}
\label{sec:disc}

In this paper, we have put forward a finite-dimensional description of large networks of globally coupled spiking neurons that are described by the 
Ermentrout-Kopell canonical model of excitable neuronal systems. Each neuron can equivalently be expressed as a phase model (theta neuron) or as a quadratic integrate-and-fire (QIF) neuron with threshold and reset going to $\pm\infty$, respectively.
In the presence of heterogeneous input currents (or natural frequencies) and/or Cauchy white noise, our formalism is exact in the thermodynamic limit.
The derivation of the set(s) of three complex ordinary differential equations, \eqref{eq:Qys} or \eqref{eq:ndv}, does not rely on assumptions of weak coupling, separation of time scales, averaging, or any other approximation. Rather, the assumptions underlying the validity of the low-dimensional description are that (i) the neurons are all-to-all connected, (ii) noise is Cauchy (and not Gaussian), and (iii) inputs are distributed according to a Cauchy-Lorentz distribution; we will comment on these assumptions further below. 
We note that in the finite-dimensional reduction, both situations---neurons are subject to Cauchy white noise, or they receive Cauchy-Lorentz distributed (time-independent) inputs---result in identical mean-field dynamics.
Yet, only in the former case there is a simple unique correspondence between the mean-field dynamics via the order parameters and the phase (voltage) distribution of theta (QIF) neurons. 
In the case of heterogeneity, one can calculate the order parameters from the distribution of phases only under certain analyticity assumptions, e.g., that the density admits an analytic continuation in the upper complex plane of inputs (or frequencies in the phase description).
Hence, the results of this paper are fully applicable to noisy ensembles, but some approaches (e.g., the resetting example) are not suitable for neurons with distributed inputs/heterogeneous frequencies. \new{At this point we also mention that while traditionally applicability of the OA reduction was restricted to the case of quenched Cauchy-Lorentz distribution of inputs (frequencies) in the classical Kuramoto setup and in the QIF model~\cite{Ott-Antonsen-08,Luke-Barreto-So-13,So-Luke-Barreto-14,Laing-14,montbrio_pazo_roxin_2015,laing_2015}, only recently it has been realized that
the same equations are valid for a population driven by independent Cauchy white noises. First it 
has been demonstrated for the Kuramoto model in \cite{Toenjes-Pikovsky-20,Tanaka-20}, and recently
extended to a QIF setup \cite{clusella_montbrio_2022}. 
We stress that Ref.~\cite{clusella_montbrio_2022}
does not go beyond the OA ansatz, in contradistinction to the full description developed above.}

We remark that the six-dimensional dynamical reduction presents an important extension of previous 
results~\cite{Luke-Barreto-So-13,So-Luke-Barreto-14,Laing-14,montbrio_pazo_roxin_2015,laing_2015,clusella_montbrio_2022} which were restricted to the OA (Lorentzian) manifold.
According to these results, the collective dynamics can only be described in the time-asymptotic limit, $t\to\infty$, or if the initial state of the neurons is meticulously instantiated. By contrast, our approach allows us to 
faithfully capture the network dynamics from arbitrary initial conditions. 
Furthermore, we can track how the collective dynamics is eventually attracted to the two-dimensional, attractive OA manifold. 
We have derived the global stability of the OA manifold (in the weak sense), whose attractiveness has already been argued in the literature~\cite{Ott-Antonsen-09,ott2010comment,pietras_daffertshofer_2016,engelbrecht_mirollo_2020,cestnik_pikovsky_2022,cestnik_pikovsky_2022a}.
In contrast to networks of conventional Kuramoto-type oscillators, where the convergence rate is given by the degree of heterogeneity and/or noise strength~\cite{cestnik_pikovsky_2022a},
here we cannot indicate an exact lower bound on the convergence rate towards the OA manifold; note, however, that here we do not consider conventional Kuramoto-type oscillators with additive noise and identical forcing fields $h_j\neq h$, but rather parameter-dependent oscillatory systems with multiplicative noise and oscillator-dependent natural frequencies $\omega_j\neq\omega_k$ and forcing fields $h_j\neq h_k$.

Our theory is valid for any mean-field coupling. As particular examples we considered all-to-all coupling via instantaneous chemical synapses as well as via electrical synapses through so-called gap junctions.
In Section~\ref{sec:obs} we provide a general framework how chemical interactions via a variety of pulses, emitted from the pre- to the postsynaptic neuron, can be incorporated through mean-field variables in our low-dimensional description. 
While we have focused here on symmetric pulse profiles, in future work we will investigate the effect of asymmetric pulses on the collective dynamics of spiking neurons.
 
As to the specific noisy and quenched inputs, we remark that in case of Gaussian white noise additional terms appear in the equations for the order parameters \eqref{eq:ni}, which does not allow for truncating the infinite system~\eqref{eq:bet} for the $\beta_n$.  Nevertheless, truncation might yield an approximative finite-dimensional description similar to that proposed in the Supplementary Material of~\cite{cestnik_pikovsky_2022} and in \cite{tyulkina_goldobin_klimenko_pikovsky_2018,Goldobin_etal-21}; this is a subject of a forthcoming research 
\new{and of particular relevance when endogenous fluctuations, e.g., in networks with sparse synaptic coupling, can be described by an effective Gaussian noise~\cite{di2022coherent}.}

Relaxing the nature of quenched heterogeneity appears straightforward, in particular if the input parameters $\eta$ are drawn from a non-singular distribution with a finite set of poles outside the real axis, see e.g., Eq.~(2) in \cite{lafuerza2010nonuniversal}. The proposed distribution allows one to approximate both Gaussian \cite{Pyragas-Pyragas-22} as well as uniform heterogeneity \cite{different_frequency_distributions,pietras2018first} with a finite-dimensional extension of the OA dynamics to arbitrary accuracy. We leave the corresponding extension of our exact low-dimensional description of spiking neurons subject to heterogeneous inputs with $q$-Gaussian or rational distributions, respectively, for future work.
As a side note, we advise caution when dealing with distributions of inputs that do not comply with the analyticity assumptions mentioned above. In those cases, transient dynamics can become nontrivial and the basins of attraction of time-asymptotic solutions depend on the choice of initial conditions and cannot be treated within the theory presented here, see, e.g., \cite{ott_platig_et_al_2008,pikovsky_rosenblum_2011}.

Finally, we mention that our approach will hold for more general networks of spiking neurons.
Possible extensions include networks with distributed synaptic weights $\kappa$, in addition to distributed inputs $\eta$, or interacting populations of excitatory and inhibitory neurons. Moreover, synapses can follow more complex synaptic kinetics and can be modeled as conductances with reversal potentials, that can also be distributed.
In these cases, it may be more important to faithfully capture transient dynamics off the OA manifold, which can readily be achieved with our exact low-dimensional description.

\acknowledgments
We thank R. Toenjes and P. Clusella for useful discussions.
BP has received funding from the European Union’s Horizon 2020 research and innovation programme under the Marie Skodowska-Curie grant agreement No 101032806.
AP and RC were supported by the DFG (Grant PI 220/21-1).


%

\end{document}